\documentclass[a4paper,11pt]{article}

\pdfoutput=1
\usepackage{jcappub}
\usepackage{graphicx}

\usepackage{amsmath}
\usepackage{array}

\newcommand{\nbox}{{\,\lower0.9pt\vbox{\hrule \hbox{\vrule height 0.2 cm
\hskip 0.2 cm \vrule height 0.2 cm}\hrule}\,}}

\newcommand{\sigmaSD}{\sigma_{\rm SD}}

\newcommand{\gev}{\text{GeV}}

\newcommand{\pb}{\text{pb}}

\newcommand{\cm}{\text{cm}}
\newcommand{\m}{\text{m}}
\newcommand{\km}{\text{km}}

\newcommand{\g}{\text{g}}
\newcommand{\s}{\text{s}}

\newcommand{\be}{\begin{equation}}
\newcommand{\ee}{\end{equation}}
\newcommand{\bea}{\begin{eqnarray}}
\newcommand{\eea}{\end{eqnarray}}
\newcommand{\baln}{\begin{align}}
\newcommand{\ealn}{\end{align}}
\newcommand{\lsim}{\lower.7ex\hbox{$\;\stackrel{\textstyle<}{\sim}\;$}}
\newcommand{\gsim}{\lower.7ex\hbox{$\;\stackrel{\textstyle>}{\sim}\;$}}

\def\parenbar#1{{\null\!                        
   \mathop#1\limits^{\hbox{\scalebox{.3}{\bf (---)}}}       	
   \!\null}}                                    

\usepackage{graphicx}

\title{\boldmath Searching for Dark Matter Annihilation to Monoenergetic Neutrinos with Liquid Scintillation Detectors}

\author[a,1]{J.~Kumar}
\author[b,2]{P.~Sandick}

\affiliation[a]{Department of Physics and Astronomy, University of Hawai'i, Honolulu, HI 96822, USA}
\affiliation[b]{Department of Physics and Astronomy, University of Utah, Salt Lake City, UT  84112, USA}

\emailAdd{jkumar@phys.hawaii.edu}
\emailAdd{sandick@physics.utah.edu}

\abstract{
We consider searches for dark matter annihilation to monoenergetic neutrinos in the core
of the Sun.  We find that liquid scintillation neutrino detectors have enhanced sensitivity
to this class of dark matter models, due to the energy and angular resolution possible
for electron neutrinos and antineutrinos that scatter via charged-current interactions.
In particular we find that KamLAND, utilizing existing data, could provide better sensitivity to such models
than any current direct detection experiment for $m_X \lesssim 15~\gev$.  KamLAND's sensitivity
is signal-limited, and future liquid scintillation or liquid argon detectors with similar
energy and angular resolution, but with larger exposure, will provide significantly better
sensitivity.  These detectors may be particularly powerful probes of dark matter with mass $\mathcal{O}(10)$ GeV.
}

\begin{document}
\maketitle
\flushbottom

\section{Introduction} \label{sec:intro}

If the dark matter-nucleon scattering matrix element is spin-dependent, then one of the leading
detection strategies is to search for the neutrinos that arise from the annihilation
of dark matter that has collected in the Sun~\cite{Silk:1985ax, Press:1985ug, Krauss:1985ks}.  Some types of neutrino detectors, such
as liquid scintillation or liquid argon detectors, can provide very good energy resolution
for an electron \mbox{(anti)neutrino.}  In this work, we describe the enhanced sensitivity of these detectors
to dark matter candidates that annihilate to monoenergetic neutrinos, and in particular
determine the current limits that can be placed by the Kamioka Liquid Scintillator Antineutrino Detector (KamLAND) with data that has already been taken.

WIMP annihilation directly to neutrino-antineutrino final states, $XX \rightarrow \nu \bar\nu$, is typically considered to be disfavored.
Indeed, if the dark matter candidate is a Majorana fermion, as is found in the MSSM for neutralino dark matter, direct annihilation to $\nu \bar\nu$ is strongly suppressed:
The initial state consists of two identical fermions, so its wavefunction must be totally antisymmetric,
implying that $s$-wave annihilation can only occur from a $J=0$ initial state.  For a $J=0$ $\nu \bar\nu$
final state, the neutrino and antineutrino must have the same helicity, so one would expect
this amplitude to be suppressed by the neutrino mass, assuming minimal flavor violation.  This
result relies on the assumption that dark matter is its own anti-particle and also on the assumption of minimal flavor violation;
if either assumption fails, the process $XX \rightarrow \nu \bar\nu$ may, in fact, have a large branching fraction.
Explicit models for which one would expect non-negligible annihilation to monoenergetic neutrino final states have been studied in~\cite{Belotsky:2001ka, Belotsky:2002sv, Barger:2007hj, Belotsky:2008vh, Esmaili:2009ks, Esmaili:2010wa}. 

Searches for neutrinos from dark matter annihilation, in the Sun or elsewhere, are sensitive to the expected neutrino energy spectrum.
If dark matter annihilates primarily to heavy Standard Model particles, such as $W$-bosons, $b$-quarks,
or $\tau$-leptons, then the decay of these products produces a continuum neutrino spectrum over a range of energies, with a high energy cut-off near the mass of the dark matter particle.
In this case, experimental searches must consider possible excesses of events over the relevant, potentially very broad, range of energies.
If, on the other hand, dark matter annihilates directly to $\nu \bar{\nu}$ final states, the signal would manifest as monoenergetic neutrinos and antineutrinos.

The energy of an electron (anti)neutrino can be measured to high
precision by liquid scintillation detectors~\cite{Learned:2009rv,Peltoniemi:2009xx} and liquid argon detectors~\cite{Adams:2013qkq}.
A high energy electron neutrino produces an $e^\pm$ through a deep inelastic charged-current interaction, and the energy of the electron shower
as well as that of the scattered parton can be well-measured at those types of detectors.  The atmospheric neutrino
background to a search for monoenergetic lines is greatly reduced as the energy resolution of the detector is improved. Furthermore, if there are a sufficiently large number of background events, it may also be possible to characterize and subtract the atmospheric background using sideband techniques.  Thus
liquid scintillation or liquid argon detectors may provide impressive sensitivity to dark matter candidates that annihilate to monoenergetic neutrinos.

Previous studies of monoenergetic neutrino signals from dark matter annihilation in the Sun have focused on heavier WIMPs accessible to the IceCube neutrino detector.  By contrast, the detectors considered here are limited in their sensitivity to low WIMP masses only by the physics of WIMP interactions within the Sun itself (evaporation)~\cite{WIMPevaporation}, and not by a loss of sensitivity to the relevant range of neutrino energies.  The detectors considered here therefore represent an excellent complement to neutrino searches for heavier WIMPs with masses $\gtrsim 100$ GeV.

In section II, we describe the general analysis underlying a monoenergetic neutrino search
at a liquid scintillation-type detector.
In section III, we implement this analysis for the specific case of KamLAND.  We conclude
with a discussion of our results in section IV.

\section{Analysis}

We focus on a search for monoenergetic electron (anti)neutrinos arising from dark matter
annihilation, $XX \rightarrow \nu_l \bar\nu_l$, and assume the annihilation cross section is flavor-independent.
A deep inelastic charged-current interaction (for example, $\nu_e d \rightarrow
e^- u$) within the target will produce electromagnetic and hadronic showers, the energies of which can be measured quite well in
liquid scintillation or liquid argon detectors.  Moreover, the direction of the electromagnetic shower
can also be reconstructed\footnote{In a liquid scintillation detector, the timing of illumination of the various photomultiplier
tubes can be used to reconstruct the direction of the electron or positron track within the scintillator~\cite{Learned:2009rv,Peltoniemi:2009xx}.},
providing a good measurement of the outgoing electron or positron's direction.
The energy resolution and angular resolution possible at liquid scintillation-type neutrino detectors make these instruments extremely powerful tools with which to search for dark matter.

\subsection{Neutrino Flux from Dark Matter Annihilation}

In general, the differential flux of monoenergetic neutrinos produced by dark matter annihilation can be
expressed as
\bea
{d^2\Phi_{DM} \over d\Omega dE} &=& {1 \over 4\pi} B_\nu \left[ \int dr\, {d\Gamma \over dV } \right] \delta (E_\nu - m_X) ,
\label{eq:flux}\eea
where $d\Gamma / dV$ is the rate of dark matter annihilation per volume,
and $B_\nu$ is the branching fraction for dark matter to annihilate to monoenergetic neutrinos, which we
take to be flavor-independent.
The differential antineutrino flux is the same.

The Sun may be an important source of monoenergetic neutrinos from dark matter annihilations.
If dark matter in the Sun is in equilibrium, then the annihilation rate is $1/2$ the rate at
which dark matter is captured via scattering against solar nuclei (see, eg.,~\cite{Peter:2009mk}).  We assume that dark matter-nucleon scattering is
a spin-dependent contact interaction, as spin-independent scattering is tightly constrained by
direct detection experiments.  We may then express this capture rate~\cite{Gould:1987ir} as
$\Gamma_C = C_0^{SD}(m_X) \times \sigmaSD^p \times [(\rho_X / \rho_\odot) (\bar v / 270~\km / \s)^{-1}] $,
where $\sigmaSD^p$ is the dark matter-proton spin-dependent scattering cross section,
$\bar v$ is the dark matter velocity dispersion of a Maxwell-Boltzmann distribution, and
where values of $C_0^{SD}(m_X)$ may be found, for example, in~\cite{Gao:2011bq}.  We then find
\bea
\int dr\, {d\Gamma \over dV } &=& {1 \over 2 r_\oplus^2} C_0^{SD} (m_X) \sigmaSD^p
\left[\left( {\rho_X \over \rho_\odot} \right) \left({\bar v \over 270~\km / \s} \right)^{-1}  \right]
\delta (\Delta \Omega) ,
\label{eq:SunRate}
\eea
where $r_\oplus = 1~{\rm AU}$ is the Earth-Sun distance, and the $\delta$-function imposes the constraint that
neutrinos will arrive from the direction of the Sun.  If we take, as a benchmark, $\langle \sigma_A v \rangle =
1~\pb \times c$ and $m_X = 10~\gev$, then the Sun is in equilibrium if $\sigmaSD^p \gtrsim 3 \times 10^{-7}~\pb$~\cite{Kumar:2012uh}.
Comparing the result of Eq.~\ref{eq:SunRate} (assuming $\sigmaSD^p \gtrsim 3 \times 10^{-7}~\pb$) to the integrated annihilation
rate near the Galactic Center as detailed in appendix \ref{sec:appendix}, it's clear that
the neutrino flux from the Sun is a few orders of magnitude larger than that from the
Galactic Center\footnote{For $m_X \gg 10~\gev$, the neutrino flux from the Sun and from the Galactic Center both scale as
$m_X^{-2}$, so the conclusion holds for the range of dark matter masses considered here.}.  Henceforth, we will thus focus on searches for neutrinos arising from dark matter annihilation
in the Sun\footnote{Dark matter can only be captured in the Sun if it scatters against nuclei.
However, this coupling need not imply that the dark matter annihilation branching
fraction to hadronic final states is dominant.}.

The effects of oscillations, absorption, and regeneration as neutrinos propagate in the Sun, from the Sun to the Earth, and in the Earth, have been well-studied in recent years.  Here, we consider only the energy range from 5 to 100 GeV, since evaporation from the Sun becomes significant for dark matter with $m_X\lesssim 4$ GeV~\cite{WIMPevaporation} and neutrino absorption suppressed the line signal for neutrino energies $E_{\parenbar{\nu}} \gtrsim 100$ GeV~\cite{Cirelli:2010xx}.  For the range of energies we consider, although interactions in the Sun will affect
neutrino flavor oscillations, the effect on the neutrino energy spectrum will be relatively small
(see, for example, \cite{Allahverdi:2012bi}).  For a specific detector, a more detailed analysis could
include the effect on the neutrino spectrum of matter effects in the Sun and Earth (the latter depend on the
location of the detector), but for a general analysis this will not be necessary.

Finally, for the range of WIMP masses considered here, we note that the assumption of dark matter capture-annihilation equilibrium is satisfied for total annihilation cross sections within a few orders of magnitude of the thermal benchmark, $\langle \sigma_A v \rangle = 1 \pb \times c$, for spin-dependent WIMP-proton scattering cross sections of $\sigma_{SD}^p \sim 10^{-4}$ pb~\footnote{The equilibration time scales as $(\sigmaSD^p < \sigma_A v > )^{-1/2}$.}.  The total annihilation rate in the Sun is then set by $\sigma_{SD}^p$ as in Eq.~\ref{eq:SunRate}, with a fraction $B_{\nu}$ of these annihilations resulting in monoenergetic neutrinos.
Our results are therefore sensitive to the branching fraction to monoenergetic neutrino final states, but robust with respect to the overall dark matter annihilation cross section today, which may be significantly smaller than the typically-assumed thermal scale.

\subsection{Effective Area of the Detector}

Assuming that the number of protons and neutrons in the detector target are the same,
the average neutrino-nucleon deep inelastic scattering cross sections for charged-current interactions with
$E \gg \gev$ are~\cite{Edsjo:1993pb},
\bea
\sigma_{\nu N} &\sim& (6.66 \times 10^{-3} \pb) \left({E \over \gev} \right) ,
\nonumber\\
\sigma_{\bar \nu N} &\sim& (3.25 \times 10^{-3} \pb) \left({E \over \gev} \right) .
\eea
We may then express the effective area of the detector, $A_\parenbar{\nu}^{eff.}=\sigma_{\parenbar{\nu} N} \left(\frac{\rho_{det}}{m_N}\right) V$, where $\rho_{det}$ is the density of the detector material, $m_N$ is the mass of a nucleon, and $V$ is fiducial volume, for (anti)neutrino scattering as
\bea
A_\nu^{eff.} &=& 2.03 \times 10^{-3} \cm^2 \left({E \over \gev} \right) \left({\rho_{det} \over 1~\g /\cm^3 } \right)
\left({V \over 10^3~\m^3 } \right),
\nonumber\\
A_{\bar \nu}^{eff.} &=& 9.88 \times 10^{-4} \cm^2 \left({E \over \gev} \right)
\left({\rho_{det} \over 1~\g /\cm^3 } \right)
\left({V \over 10^3~\m^3 }\right).
\eea

We will only consider events in which the $e^\pm$ falls within an angle $\theta_{cone}$ of
the Sun~\cite{Jungman:1995df}, where
\bea
\theta_{cone} &=& 0.37 \sqrt{10~\gev \over E}.
\eea
For electron neutrinos (antineutrinos) arriving from the sun, $f_\nu$ ($f_{\bar \nu}$) is the fraction of electrons (positrons) produced
by charged-current interactions that fall within this cone.  $f_{\nu, \bar \nu}$ is roughly determined by the 
angular dependence of the matrix element for left-handed quarks to scatter against left-handed neutrinos and 
right-handed anti-neutrinos, respectively.  $f_{\nu, \bar \nu}$ can then be expressed as
\bea
f_\nu &=& \frac{1-\beta^2}{2}\int_{\cos\theta_{cone}}^1 dx \, \frac{1}{(1-\beta x)^2}
\nonumber\\
f_{\bar{\nu}} &=& \frac{1-\beta^2}{2}\int_{\cos\theta_{cone}}^1 dx \,  \frac{1}{(1-\beta x)^2}\frac{3}{4} (1+\cos\theta_{cone})^2,
\label{eq:f_fbar} 
\eea
where $\beta$ is the boost from center-of-mass frame to lab frame.
We find $f_\nu \approx 0.4$ and $f_{\bar \nu} \approx 0.8$, results which are mildly dependent on the (anti)neutrino energy.  In the following analysis, we use the full expression for $f_\parenbar{\nu}$ given in eq.~\ref{eq:f_fbar}. 

\subsection{Signal and Background Events}

The expected number of events in a particular detector arising from dark matter annihilation can then be expressed as
\bea
N^{\parenbar{\nu}_e}_{sun} &=& 0.68 \times T \left({B_\nu \over 3} \right)
\int_{E_{min}}^{E_{max}} dE \, {d\Phi_{DM} \over dE} \times A_{\parenbar{ \nu}}^{eff.} \times f_{\parenbar{ \nu}} ,
\eea
where $T$ is the detector runtime, $E_{min, \,max}= m_X [1 \pm (\epsilon/2)]$, and $\epsilon$ is the energy resolution of the detector.
Again, $B_\nu$ is the branching fraction for dark matter annihilation to all neutrino flavors; since we assume
that annihilation is flavor-independent, $\sim 1/3$ of the neutrinos reaching the detector will be
electron neutrinos even after including vacuum and matter oscillation effects~\cite{Lehnert:2007fv}.  The factor of $0.68$ accounts for the fact that $68\%$ of events fall within a bin of width $\epsilon$.
Similarly, the number of events due to the atmospheric neutrino background can be expressed as
\bea
N^{\parenbar{\nu}_e}_{atm} &=& T \int_{E_{min}}^{E_{max}} dE \, {d^2 \Phi_{atm}^{\parenbar{\nu}_e} \over d\Omega \, dE} \times A^{eff.}_{\parenbar{\nu}}
\times \left[(2\pi)(1-\cos \theta_{cone}) \right] ,
\eea
where ${d^2 \Phi_{atm} \over d\Omega \, dE}$ is the atmospheric electron (anti)neutrino flux.  We use the angle averaged
atmospheric electron (anti)neutrino fluxes given in~\cite{Honda:2011nf}, which, in the relevant energy range, are well-approximated by the
power law fits
\bea
{d^2 \Phi_{atm}^{\nu_e} \over d\Omega dE} &\sim&  (4.17 \times 10^{-2} \cm^{-2} \s^{-1} {\rm sr}^{-1} \gev^{-1} )
\times \left(0.80 + {E \over \gev} \right)^{-3.490},
\nonumber\\
{d^2 \Phi_{atm}^{\bar \nu_e} \over d\Omega dE} &\sim&  (2.42 \times 10^{-2} \cm^{-2} \s^{-1} {\rm sr}^{-1} \gev^{-1} )
\times \left(0.53 + {E \over \gev} \right)^{-3.417}.
\eea

Given the number of signal and background events expected at a neutrino detector, one can then
determine sensitivity of the detector to a dark matter model for any choices of
$m_X$ and $\sigmaSD^p$.

\section{Bounds from KamLAND}

In this section, we apply the analysis framework above to the specific case of the KamLAND detector and the data it has accumulated over more than
10 years.
KamLAND~\cite{Eguchi:2002dm} is a liquid scintillation neutrino detector with $\rho_{det} \approx 0.8~\g / \cm^3$ and an approximately spherical
inner detector with volume $\sim 1000~\m^3$.  We define the fiducial volume of the detector such that if a charged-current interaction produces an $e^\pm$ within the fiducial volume, at least 10 radiation lengths ($\sim 4.3~\m$) will be contained within the
inner detector.  For KamLAND, the fiducial volume is $\sim 500~\m^3$~\cite{Kumar:2012uh}.
If an electron (anti)neutrino with energy greater than $1~\gev$ participates
in a charged-current interaction within the fiducial volume, it is estimated that the energy of the original neutrino can be
reconstructed to within a few percent accuracy, and that approximately 3600 live-days of data are available
for such an analysis~\cite{KamLAND_energy_res}.
As a benchmark, we will take the energy resolution for this analysis to be $\epsilon =5\%$.
It is also estimated that the charged lepton flavor can be determined with very high efficiency, and that the
direction of the electromagnetic shower can be determined with a resolution much better than $\theta_{cone}$ for the
energy range of interest~\cite{Peltoniemi:2009xx}.

\begin{figure}[t]
  \centering
    \includegraphics[width=0.7\textwidth]{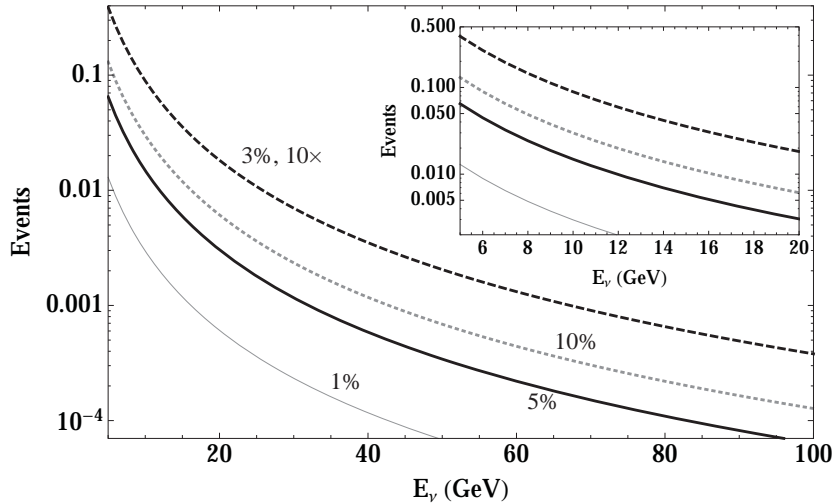}
    \caption{The expected number of background events as a function of the central bin energy for a KamLAND exposure of 3600 days with energy resolution of 1\% (solid grey), 5\% (solid black), and 10\% (dotted grey).  We also show the projected number of background events as a function of energy for a detector with an exposure 10 times larger than that of KamLAND with an energy resolution of 3\% (dashed black).  We note that the dotted grey curve also represents the expected number of background events for a detector with an exposure 10 times larger if the energy resolution is 1\%.
\label{fig:background}}
\end{figure}

We note that for the exposure of KamLAND, one expects far less than one background event within each energy
bin, whose size is determined by the energy resolution.  In Figure~\ref{fig:background}, we show the expected number of background events as a function of the central bin energy for a KamLAND exposure of 3600 days and energy resolution of 1\% (solid grey), 5\% (solid black), and 10\% (dotted grey).  We also show the projected number of background events as a function of energy for a detector with an exposure 10 times larger than that of KamLAND with an energy resolution of 3\% (dashed black).  The inset shows the low-energy regime in more detail.  It is clear that, for the exposure of KamLAND, the atmospheric neutrino background is negligible.

\begin{figure}[t]
  \centering
    \includegraphics[width=0.7\textwidth]{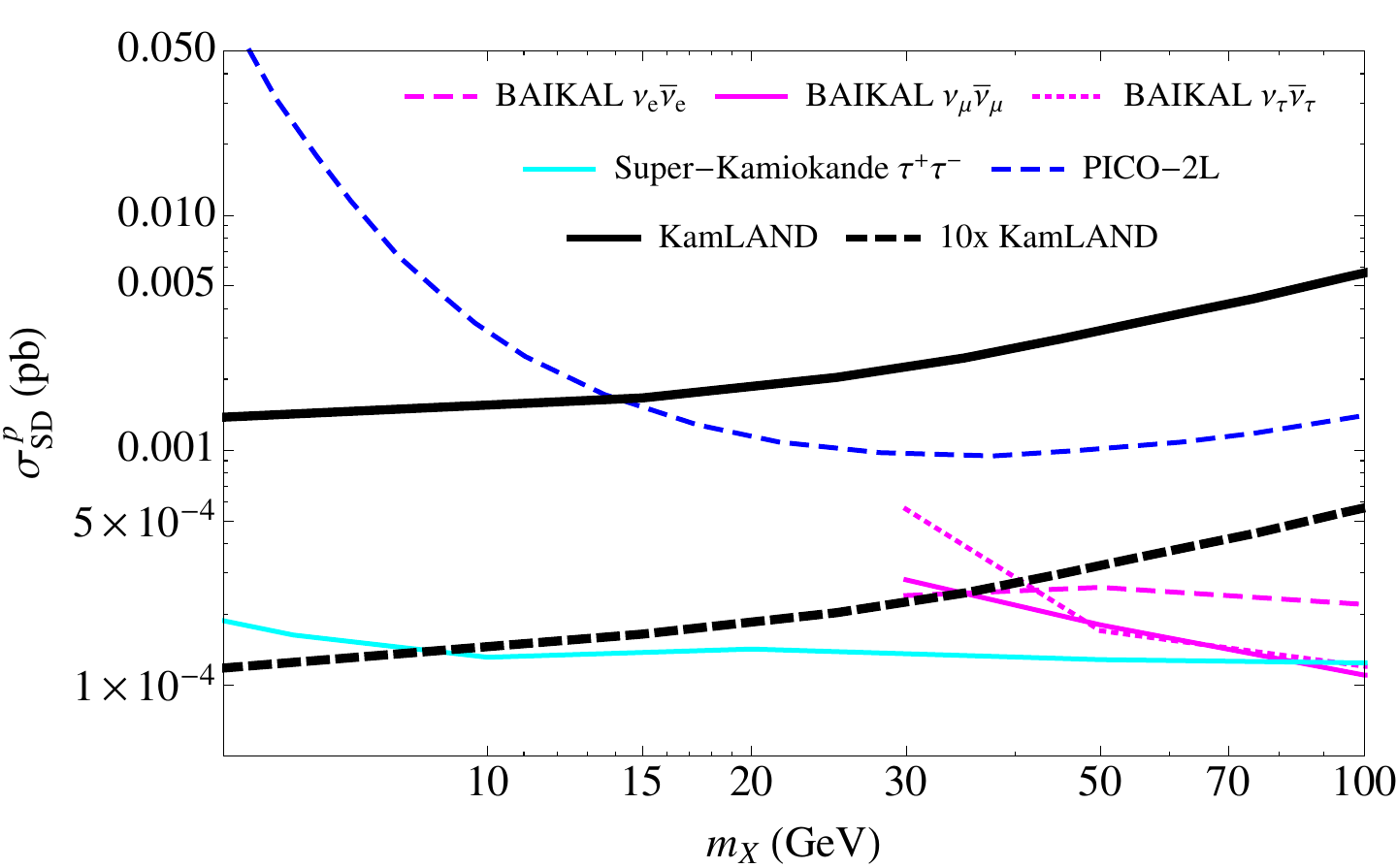}
    \caption{90\% CL upper limit on the spin-dependent dark matter-proton elastic scattering cross section as a function of dark matter mass assuming zero events at KamLAND in 3600 days of data (thick black), as well as the expected sensitivity of a liquid scintillation detector with a factor 10 larger exposure than KamLAND and with a $3\%$ energy resolution (thick black dashed).  We assume $B_\nu =1$, $\rho_X = \rho_\odot$ and $\bar v = 270~\km /\s$.
    For comparison, the dashed, solid, and dotted magenta curves are the upper limits from Baikal NT200 detector~\cite{Avrorin:2014swy}, assuming dark matter annihilation to $\bar \nu_e \nu_e$, $\bar \nu_\mu \nu_\mu$ and $\bar \nu_\tau \nu_\tau$, respectively, the solid cyan curve is the upper limit from Super-Kamiokande, assuming annihilation to $\tau^+ \tau^-$~\cite{SuperKprl}, and the dashed blue contour is the upper limit from the PICO-2L Bubble Chamber dark matter experiment~\cite{PICO2Lpub}.
\label{fig:limitplot}}
\end{figure}

A 90\% confidence level upper limit on the spin-dependent dark matter-proton elastic scattering cross section as a function of dark matter mass is
obtained (see, for example, ~\cite{Feldman:1997qc}), assuming zero observed events in each relevant energy bin during 3600 KamLAND live-days.
The expected upper limit on the spin-dependent dark matter-proton scattering cross section from KamLAND is shown in Figure~\ref{fig:limitplot} as the thick black contour for $B_{\nu}=1$, $\rho_X = \rho_\odot$ and $\bar v = 270~\km / \s$~\footnote{As the number of signal events is proportional to $\sigmaSD^p B_\nu (\rho_X / \rho_\odot)(\bar v / 270~\km /\s)^{-1}$, the sensitivity for any choice of $B_\nu$, $\rho_X$ and $\bar v$ can be determined
by a simple rescaling.}, as well as the expected sensitivity of a liquid scintillation detector with a factor 10 larger exposure than KamLAND and with a $3\%$ energy resolution (thick black dashed).
 For comparison, the dashed, solid, and dotted magenta curves are the upper limits from Baikal NT200 detector~\cite{Avrorin:2014swy}, assuming dark matter annihilation to $\bar \nu_e \nu_e$, $\bar \nu_\mu \nu_\mu$ and $\bar \nu_\tau \nu_\tau$, respectively, the solid cyan curve is the upper limit from Super-Kamiokande, assuming annihilation to $\tau^+ \tau^-$~\cite{SuperKprl}, and the dashed blue contour is the upper limit from the PICO-2L Bubble Chamber dark matter experiment~\cite{PICO2Lpub}.  For clarity, we display only the leading published results (or, in the case of PICO-2L, publicly released but not yet published), omitting competitive but subleading bounds from direct and indirect searches such as PICASSO~\cite{Archambault:2012pm}, SIMPLE~\cite{Felizardo:2011uw}, COUPP~\cite{Behnke:2012ys}, the Baksan Underground Scintillator Telescope~\cite{Boliev:2013ai}, and the IceCube Neutrino Telescope~\cite{Aartsen:2012kia}.

The estimated sensitivity is
relatively robust, even if the actual energy resolution is different from our benchmark estimate by a factor of a few.  This is clear from Fig.~\ref{fig:background} for the current KamLAND exposure of $\sim 3600$ days.  Even for an exposure 10 times larger, the atmospheric background only
approaches $\mathcal{O}(1)$ event in an energy bin for the very lowest energies considered.
For $\epsilon = 3\%$, the expected number of background events in an energy bin is never larger than 0.39, for $E_\nu = 5$ GeV, and falls to less than 0.09 for $E_\nu \geq 10$ GeV.  If a single event is observed in an energy bin, the actual sensitivity will drop by $\sim 40-50\%$.

As is clear from Figure~\ref{fig:limitplot}, at this moment, the limit obtainable from KamLAND data would be stronger than limits from traditional direct dark matter searches (i.e.~for dark matter collisions with nuclei in terrestrial detectors) for WIMP masses $\lesssim 15$ GeV, assuming dark matter annihilates to $\nu \bar{\nu}$ final states with $B_\nu=1$.  The sensitivities of neutrino telescopes such as Super-Kamiokande, Baksan, and IceCube are specific to the dark matter annihilation final state, though harder neutrino spectra generally yield stronger constraints on the scattering cross section.  For those detectors,
actual limits for monoenergetic neutrino final states are likely stronger than the limits shown here.  We encourage collaborations to perform these searches.

Finally, we note that KamLAND's sensitivity is signal-limited, due to its relatively small exposure compared to that of much larger water Cherenkov
detectors such as Super-Kamiokande.  As such, the utility of liquid scintillation detectors for searches for dark matter annihilation to
monoenergetic neutrinos can only be fully realized for detectors with a much larger exposure.  For example, for an exposure ten
times larger than our benchmark KamLAND exposure, cross sections as small as $\sigma_{SD}^p \approx 10^{-4} \,(6\times10^{-4})$ pb could be probed
for $m_X =10 \,(100)$ GeV.  Such an experiment would be an very powerful probe of the interactions of dark matter with nuclei in the Sun.  This is especially true at low WIMP masses, where liquid scintillation detectors have robust sensitivity to the WIMP-proton scattering cross section down to $m_X \approx 4$ GeV, below which evaporation from the Sun becomes significant.

\section{Conclusions}

In this work, we have considered the sensitivity of liquid scintillation neutrino
detectors to dark matter annihilation in the Sun in the case that the dark matter annihilates to
monoenergetic neutrinos.  Because liquid scintillation detectors can reconstruct with
good accuracy all of the energy of an electron (anti)neutrino interacting through a charged-current
interaction, this is essentially a search for an excess in a single energy bin.  The
resulting reduction in background greatly enhances the sensitivity of liquid scintillation
neutrino detectors.  As an example, we have considered the sensitivity that KamLAND may be
capable of, using existing data.

As may be expected, the sensitivity of liquid scintillation detectors found here exceeds those of other estimates~\cite{LS_Detector_DM_Search},
which focussed on annihilation to final states that included a continuum spectrum of neutrinos, in which case a signal would emerge over a broad range of neutrino energies rather than within a single energy bin.  With its current
exposure, KamLAND would expect far less than one background event in any single energy bin for energies $\gtrsim5$ GeV.
In fact, for any liquid scintillation detector, the expected sensitivity will increase linearly with exposure in this low-background regime
(this low-background regime should extend up to an exposure of roughly 10 times the current exposure of KamLAND for all WIMP masses considered, and to considerably larger exposure for $m_X \gtrsim 100$ GeV).  If the exposure becomes large enough that the expected number of background events in an energy bin approaches one, the assumption of zero events in any bin is, of course, no longer valid for estimating the sensitivity.

It is worth noting that liquid argon-based neutrino detectors can also reconstruct the energy and direction of
an electron shower created by a charged-current interaction.  A similar analysis could thus be performed
for such detectors.  Moreover, future liquid argon-based detectors may have a much larger exposure than KamLAND.  For
example, the LBNE far detector is foreseen to be a 34 kiloton liquid argon time projection chamber~\cite{Adams:2013qkq}.
If the energy resolution is as good as a few percent, then such detectors
could obtain competitive sensitivity, placing them among the leading experimental searches for particle dark matter.

\vskip .2in
\textbf{Acknowledgments}

We are grateful to K.~Choi, C.~Kelso, J.~G.~Learned and M.~Sakai  for useful discussions.  The work of J.~Kumar is
supported in part by NSF CAREER Grant No.~PHY-1250573. The work of P. Sandick is supported in part by NSF Grant No.~PHY-1417367.
J.~Kumar and P.~Sandick would like to thank the University of Utah and University of Hawai'i, respectively,
for their hospitality and partial support during the completion of this work.

\appendix
\section{Monoenergetic Neutrinos from the Galactic Center}
\label{sec:appendix}


Indirect signals of dark matter annihilation in most astrophysical bodies do not rely on dark matter scattering with nuclei in order to become gravitationally bound, as is the case for the annihilation signal from the Sun.  Instead, the annihilation rate is proportional to the annihilation cross section, rather than the dark matter-nucleon scattering cross section.
For dark matter annihilation in the Galactic Center region, the integrated annihilation rate may be expressed as
\bea
\int dr\, {d\Gamma \over dV } &=& {\langle \sigma_A v \rangle \over 2m_X^2} (r_\odot \rho_\odot^2) J ,
\label{eq:GC}\eea
where $\langle \sigma_A v\rangle$ is the total dark matter annihilation cross section times velocity, $r_\odot = 8.33~{\rm kpc}$ is the distance from the earth to the Galactic Center~\cite{Gillessen:2008qv}, and
$\rho_\odot = 0.3~\gev /\cm^3$ is the putative density of dark matter in the solar system~\cite{Bovy:2012tw}.  The quantities $r_\odot$ and $\rho_\odot$ serve to normalize the $J$-factor, which encodes the relevant information regarding the dark matter density along the line of
sight to the target.  Typical $J$-factors for the Galactic Center are ${\cal O}(1-10)$~\cite{Cirelli:2010xx}.  We have
assumed that the dark matter particle is its own anti-particle; if the particle and anti-particle are distinct, this
annihilation rate is reduced by a factor of 2.  The differential flux of monoenergetic neutrinos produced by dark matter annihilation in the Galactic Center (or other region, given the appropriate $J$-factor) may be calculated according to equation~\ref{eq:flux} with the integrated annihilation rate as in~\ref{eq:GC}.

\end{document}